\documentclass{article}
\usepackage{spconf, amsmath,graphicx,url}

\title{Building A modal-balanced BlockChain with Semantic Reconstruction}

\name{Zhijie Tan$^{\dag}$, Xiang Yuan$^{\dag}$, Shengwei Meng$^{\ddag}$, Yakun Huang$^{\ddag}$, Weiping Li$^{\dag}$, Zhonghai Wu$^{\dag}$, Tong Mo$^{\dag}$}
\address{$^{\dag}$School of Software and Microelectronics, Peking University, Beijing, China \\ $^{\ddag}$School of Computer Science, Beijing University of Posts and Telecommunications, Beijing, China}

\begin{document}

\maketitle

\begin{abstract}
The current large blockchain systems (BTC Lightning network, Ethereum, etc.) are generally facing the problems of low persistence rates and high storage costs. Therefore, users tend to store single modal (textual) information on the existing blockchain systems. Inspired by semantic communication algorithms, this paper presents a new algorithm to solve the serious imbalance between textual and visual modals on blockchains. After semantic sampling of the original visual image, the resulting semantic text will be stored on the chain, and the end users can reconstruct a semantically similar image using the \textbf{R}elative \textbf{O}ptimal \textbf{S}emantic \textbf{I}sotope \textbf{S}election algorithm. Experiments on the DIV2K dataset show that the blockchain with our algorithm can achieve 430,000 times the storage capacity and 550,000 times the persistence rate for the original visual data with acceptable semantic information loss.
\end{abstract}
\begin{keywords}
multimodal, blockchain, semantic reconstruction
\end{keywords}
\section{Introduction}
\label{sec:intro}

The blockchain is promised to be the fundamental infrastructure of the Metaverse with the features of permanent file storage and a decentralized verification mechanism. However, these features implied the curses to the capacities of file size and persistence rate when one file is going to be up-chained. The permanent file storage feature implied that the total disk size of the blockchain is monotonically increasing while the decentralized verification implied the file persistence should be verified in more than 50\% verification nodes. Even worse, such curses are growing up with the node number. Take Ethereum \cite{dannen2017introducing}, which currently has nearly 400,000 active addresses, as an example: From September 15th, 2021 to September 15th, 2022, the average block generation time is between 12 and 14 seconds, and the average block size is about 2 MB\footnote{Data from \url{https://www.ycharts.com/}.}. Such persistence bandwidth (about 10MB/min) is extremely limited for visual modal data which will cause severe congestion on the blockchain. Furthermore, the large-scale visual data persistence will bring unacceptable storage pressure for every decentralized node.  

To solve the problem of such curses, it is common to use IPFS (InterPlanetary File System) \cite{daniel2022ipfs} or other cloud storage methods (Google Drive, etc.) to generate a unique hash or hyperlink for the corresponding visual data. However, the eternal storage of external file links does not claim the permanent storage of data: The accessibility and existence of files in the IPFS or other cloud disks are not guaranteed by the blockchain itself, which sources the new uncertainty into the design of the blockchain systems and is therefore not a perfect solution. Besides using the external file links, traditional image compression methods (500 compressed images whose compression ratio is set to extremely high to 1:100 in one minute) still have great difficulty achieving the ideal persistence rate target to be up-chained considering a communication network consisting of nearly 400,000 users. Therefore, while the current mobile Web2 world is dominated by visual data in social media which is easier to understand for every person compared to the textual data, the blockchain for the future Web3 world still suffers the extremely rare visual data. 

Inspired by current popular semantic communication algorithms\cite{huang2021aitransfer, xie2021deep}, the core idea of this paper is swapping off-chained computing resources for up-chained persistence rate and file size. For blockchains with decentralized nodes, off-chained computing tasks can be completed parallelly while up-chained data persistence needs to be verified in serialized blockchain blocks. The effect of off-chained computing tasks will be weakened when the blockchain node number grows. Thanks to the breakthroughs in the fields of the image-to-text \cite{li2022blip, stefanini2022show}, text-to-image \cite{rombach2022high, ramesh2022hierarchical} and other multimodal technology \cite{radford2021learning}, a powerful pre-training model can extract the high-level semantic information of visual images and another model can reconstruct \textbf{semantic isotopes} (different reconstructed images generated from the same semantic information). It should be emphasized that compared to the image compression method, our semantic reconstruction algorithm pays more attention to semantic similarity instead of pixel-to-pixel visual similarity. Human beings see an image and they understand the high-level semantic information of this image. If this image is slightly shifted or rotated, this would not have a huge effect on the semantic information but will completely destruct the common visual similarity evaluation indexes like PSNR (Peak Signal-to-Noise Ratio) \cite{sheikh2006statistical} or SSIM (Structural SIMilarity) \cite{wang2004image}. To address the challenges of the curses, this paper presents an algorithm where a blockchain requires only the corresponding natural semantic text ($T$) to be up-chained. However, during the process of semantic reconstruction, there exists a problem that reconstructing the high dimensional visual data from the low dimensional textual data will inevitably bring some randomness. The randomness will be reduced when the certainty of the semantic prompts increases. 
% A direct method to increase the certainty is using models with more semantic understanding to the images or in other words using more intelligent models which will capture and reconstruct the images with less semantic loss. 
This paper designs the \textbf{R}elative \textbf{O}ptimal \textbf{S}emantic \textbf{I}sotope \textbf{S}election (\textbf{ROSIS}) algorithm to reduce the randomness which is model-independent.

% \begin{figure*}
% \centering
% \includegraphics[width=0.80\textwidth]{model.png}
% \caption{Overall architecture of the semantic reconstruction framework. Phase 1: The semantic sampling phase. The raw image is converted to a semantic text T. Phase 2: The ROSIS phase. N semantic isotopes, $I_{1}$, .., $I_{n}$ are generated by Text2Image model then converted to $T_{1}$, .., $T_{n}$. The semantic distance between $T_{s}$ and $T$ will be calculated and  $T_{s}$ with the minimum distance will be selected as $T_{rec}$. \textless$T_{rec}$, $s_{rec}$\textgreater will be uploaded to the blockchain. Phase 3: The semantic reconstruction phase. The blockchian nodes can use \textless$T_{rec}$, $s_{rec}$\textgreater to reconstruct the semantic isotope $I_{rec}$.} \label{fig1}
% \vspace{-1.5em}
% \end{figure*}

We summarize the main contributions as follows: A semantic reconstruction method is formulated to bring more visual data to be up-chained. The ROSIS algorithm is designed to reduce semantic loss during semantic reconstruction. Experiments show the great potential of this method.

\section{Methodology}
\label{sec:format}
The prototype framework of our algorithm is shown in Fig. 1. The whole algorithm consists of a semantic sampler (Image2Text model), a semantic reconstructor (Text2Image model), and the ROSIS algorithm. This section will concentrate on the definitions and designs of our algorithm. 

\subsection{Problem Formulation}
The learning-based image reconstruction process can be defined as:
\begin{equation}
    I_{o} \stackrel{f(\theta_{1})}{\longrightarrow} X  \stackrel{g(\theta_{2})}{\longrightarrow} I_{r}
\end{equation}

In (1), $I_{o}$ is the original  image, $f(\theta_{1})$ is a sampling mapping with parameters $\theta_{1}$, $X$ is the intermediate representation, $g(\theta_{2})$ is a reconstruction mapping with parameters $\theta_{2}$ and $I_{r}$ is the reconstructed image. From the perspective of information bottleneck theory\cite{tishby2000information, shwartz2017opening}, $f(\theta_{1})$  aims to minimize the mutual information $MI(I_{o},X;\theta_{1})$ while $g(\theta_{2})$ aims to maximize $MI(X,I_{l};\theta_{2})$. $I_{l}$ is the lossy compressed image downsampled from $I_{o}$. Therefore, the image lossy reconstruction problem can be defined as:
\begin{equation}
    \mathop{\arg\min}\limits_{\theta_{1}, \theta_{2}} MI(I_{o},X;\theta_{1}) - \gamma MI(X, I_{l};\theta_{2})
\end{equation}
$\gamma$ is used to balance $f$ and $g$. 
And the semantic reconstruction problem can be similarly defined as:
\begin{equation}
    \mathop{\arg\min}\limits_{\theta_{1}, \theta_{2}} MI(Inf(I_{o}),T;\theta_{1}) -  \gamma MI(T,Inf(I_{l});\theta_{2})
\end{equation}
In (3), $Inf(I)$ is the semantic representation of $I$ , and $T$ is an intermediate semantic representation. Our algorithm chooses text (natural language) as the semantic representation.

\subsection{Decoupled Semantic Sampler and Reconstructor}
% For the blockchain, a system with extremely limited file storage and bandwith, the traditional image reconstruction methods seems far behind. As the introduction states, the Ethereum running at 10 MB/min. Considering the 400,000 users, each user can enjoy the blockchain at 1/40,000 MB/min.  In contrast, the family user in China can enjoy the Internet at 2,757 MB/min \footnote{\url{http://www.ciita.org.cn/news/2832.html}}. The huge gap of magnitude calls for a new reconstruction method which can complete the breakthrough of image compression ratio.
There will be further doubt about why $X$ needs to be a text in (3). From the aspect of compression ratio, $X$ can be compressed into a lighter representation. However, such representation brings a hidden assumption that $g$ can map every $x\in X$ to an acceptable $i_{r}\in I_{r}$, which brings a restriction: $f$ and $g$ is tightly coupled (usually trained on the same dataset or for a specific task).  This restriction will require that all of the sampling and reconstruction models be up-chained. If not up-chained, $x$ without semantic information cannot be decoded into an acceptable image. Then the permanent file storage feature will be unsatisfied. Therefore, the user needs to download the corresponding reconstructor model from the blockchain, which will bring heavy communication and storage pressure for the blockchain. 
% For $f$ and $g$ which are need to be up-chained, mark the learning-based $f$ and $g$ with the corresponding parameter ($\theta1$ and $\theta2$) as $f(\theta1$) and $g(\theta2$). Then (2) will be extended as:
% \begin{equation}
%     \mathop{\arg\min}\limits_{X, \theta1, \theta2} MI(I_{ori};X) - MI(X;I_{ori})
% \end{equation}
% (4) can be understood as: $f(\theta1)$ and $g(\theta2)$ is the best coding algorithm (to get the best $X$) with the best parameters. When the result of (4) is the same, the less parameters mean better.
If $x$ is a semantic text, $f$ and $g$ can be any pair of powerful image-to-text and text-to-image pretrained models. There will be no extra communication or storage pressure for the blockchain. As a result, our method adopts a decoupled semantic sampler and reconstructor, which means these models and images are transparent to both sides. Thus, $\theta_{1}$ and $\theta_{2}$ of $f$ and $g$ can hardly be finetuned. In fact, $\theta_{1}$ and $\theta_{2}$ are fixed in our proposed algorithm.

\subsection{The Relative Optimal Semantic Isotope Selection (ROSIS) Algorithm}
\label{sec2.3}
% For the text-to-image algorithm used in 2.2, different random seeds will generate different images. Thus, it is very important to choose a seed that can produce an optimal semantic isotope $I_{rec}$  to reduce the semantic loss. 
% For the calculation of image similarity, SSIM and PSNR are common metrics. However, these methods are not very satisfactory: PSNR and SSIM pay more attention to the pixel-to-pixel (or patch-to-patch) similarity of visual objective features (brightness, structure, etc.) and cannot estimate the semantic subjective similarity that needs to be understood. Semantic similarity can better describe subjective human understanding of image information in a lower text dimensions compared to vision dimensions, which is more computing-friendly and 
% explainable especially for the current deep learning models.
Because $\theta_{1}$ is fixed, $T$ becomes a fixed value. The fixed $T$ states that the current problem is slightly different from (3) and mutual information can be replaced with cosine distance.
The semantic reconstruction problem in (3) can be converted as:
\begin{equation}
    \mathop{\arg\min}\limits_{\hat{T}} Cosine\_dist(T, Inf(\hat{I}_{r}))
    % \vspace{-0.5em}
\end{equation}

$\hat{I}_{r}$ is the reconstructed image from $\hat{T}$. $\hat{T}$ is the combination of the original $T$ and a random seed. Since the original image and the sampler is invisible to the reconstructor caused by the transparency mentioned in section 2.2, the text $T$ is the only reliable signal in (4) while in (2) and (3) $I_{o}$ is more reliable. With a compression ratio of 10000:1 or higher, reconstructing $i_{r} \in I_{r}$ from $t \in T$ will be born with uncertainty for the general task. It appears as for the Text2Image models, different random seeds will bring different images even with the same input text $T$. So we propose the ROSIS algorithm: generate different random seeds to get different $\hat{T}$ and choose the semantic closest image as the final $I_{r}$. For the tasks whose datasets and models are known beforehand, the ROSIS algorithm can be applied as a model and data independent method to acquire extra information gain based on the finetuned models. For example, the face-to-id \cite{zhao2003face} and id-to-face \cite{sun2022anyface} tasks can be improved by the ROSIS algorithm. If there are some similar faces in the dataset, the id-to-face model may generate some similar faces. Then the best face can be chosen.

% In (1), $inf(I)$ is a vector. It means that firstly $BLIP$ in 2.1 will be deployed to convert the image $I$ into a text prompt $T$, and then $T$ will be encoded by $all$-$MiniLM$-$L6$-$v2$ \cite{wang2020minilm, reimers-2019-sentence-bert} into $inf(I)$. $cos\_dist$ in (1) means the cosine distance between two vectors $u$ and $v$, which can be calculated by:
% \begin{equation}
%     cos\_dist (u, v) = \frac{{\left\|u \right\|}_{2}{\left\|v \right\|}_{2}-u^{T}v}{{\left\|u \right\|}_{2}{\left\|v \right\|}_{2}}
% \end{equation}

% In (2), ${\left\|\cdot \right\|}_{2}$ denotes the $l2$ normalization. 

\begin{figure*}
\centering
\includegraphics[width=0.93\textwidth]{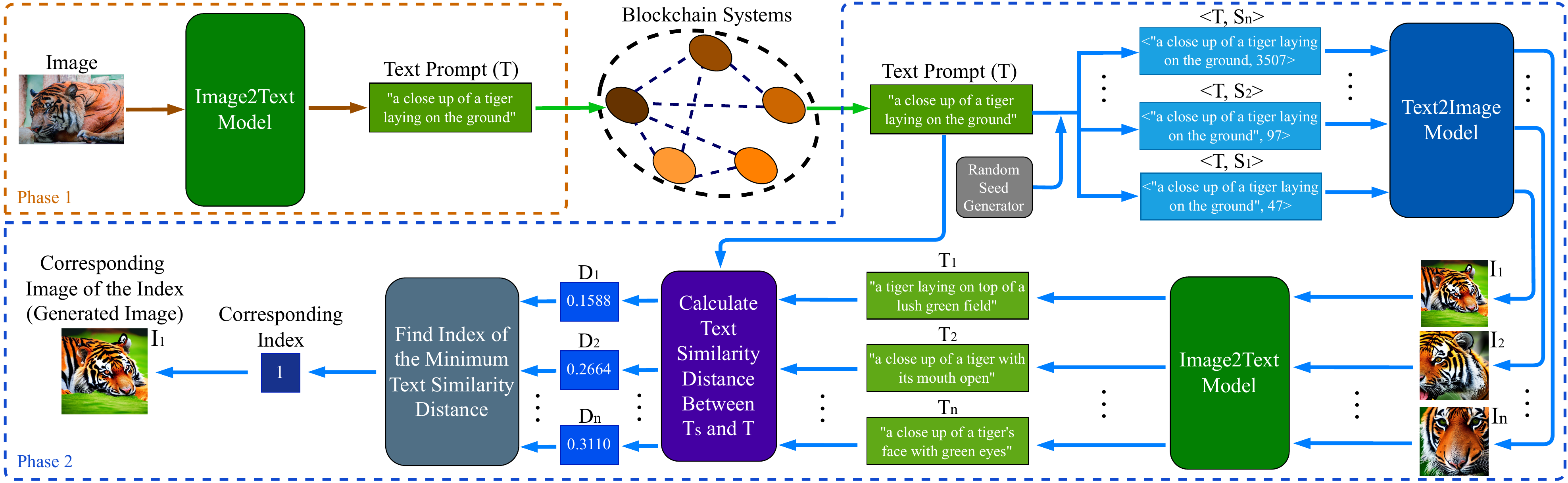}
\caption{ The overall architecture of the semantic reconstruction prototype. Phase 1: the semantic sampling phase. The raw image is converted to a text prompt $T$. Phase 2: the semantic reconstruction phase. The text prompt $T$ is fetched from the blockchain. We generate $n$ semantic isotopes, $I_{1}$, .., $I_{n}$ from $T$ by Text2Image model, then converted them to $T_{1}$, .., $T_{n}$ separately. The semantic distance between $T_{s}$ and $T$ will be calculated and $I_{r}$ with the corresponding index of the minimum semantic distance will be considered as the final reconstruction result.} \label{fig1}
\vspace{-1.5em}
\end{figure*}

\vspace{-1em}
\section{Experiments}
\vspace{-0.5em}
\subsection{Experiment Environment \& Dataset \& Metrics}

\textbf{Experiment Environment}: The operating system used in the experiment part is ubuntu 20.04, the CPU is Intel Xeon (R) E5-2603 v4@1.7GHz, and the memory size is 64GB. In this paper, PlatONE\footnote{\url{https://github.com/PlatONEnterprise/PlatONE-Go}} is used as a blockchain system for simulation. The maximum gas limit is 150m, and the difficulty is 1. 10 nodes in the blockchain and 4 random seeds for the ROSIS algorithm. We use 4 Nvidia V100 GPUs for the deep learning models. The image is first converted to the serialized data in base64 format and then partitioned per 100KB. 

\textbf{Dataset}: The dataset used in this article is DIV2K\cite{agustsson2017ntire}. DIV2K is composed of 1000 high-resolution (2k) images, containing a large number of images with different contents (humans, tigers, buildings, etc.), which can effectively verify the semantic reconstruction performance for different objects. 

% At the same time, images are compressed using common lossy formats such as .webp and .jpg. The W.75/J.75 in Table 1 means that the original image (.png) in DIV2K is saved as .webp/.jpg using the default quality 75 with Pillow. The W.1/J.1 in Table 1 means that the original image (.png) in div2k is saved as .webp/.jpg using quality 1 with Python Pillow. The default quality parameter of 75 can preserve image details well and save storage space with a 10-12x (times) compression ratio. When the quality parameter is 1, the image loses a lot of detail, and there is significant blurring and smearing. An example of a high-resolution (2040*1536) image, its compressed results, and its semantic isotope are shown in Fig. 2.  

\textbf{Metrics}: \textbf{TPT (Total Persistence Time)}: the total persistence time after all files are transferred to the blockchain and persisted on-chained successfully. The start time is the time that the command is sent to start uploading the image, and the end time is the generation time of the last block containing the image data.  \textbf{WT (Waiting Time)}: total time consumed by an image from the command to start uploading to the generation of the last block containing the image data. Limited by the block size of the blockchain, an image often needs to be divided into different blocks and then uploaded to the blockchain. The waiting time reflects the real-time congestion situation of the blockchain. \textbf{mWT (minimum Waiting Time)/ MWT (Maximum Waiting Time)}: the time required for the fastest/slowest completion of being up-chained among all the images. They reflect the waiting time that the user should suffer under the optimal/worst situation. \textbf{F.S. (File Size)}: total storage space for all files. \textbf{Dis. (Distance)}: the cosine distance between $Inf(I_ori)$ and $Inf(I_rec)$. This value is between 0 and 2. \textbf{TSS (Time of Semantic Sampling)/ TRS (Time of Semantic Reconstruction)}: the time for semantic sampling and semantic reconstruction for one image. All of the above metrics follow the rule of the smaller the better. 

% \begin{figure}[htb]

% \begin{minipage}[b]{1.0\linewidth}
%   \centering
%   \centerline{\includegraphics[width=8.5cm]{fig3.png}}
% \end{minipage}
% \caption{From \textbf{(a)} to \textbf{(f)} is in the turn of original image, .jpg with quality 75, .webp with quality 75, semantic isotope, .jpg with quality 1, .webp with quality 1.}
% %
% \vspace{-2em}
% \end{figure}

\subsection{Algorithm Performance on F.S. and P.T.}
\vspace{-0.8em}
A prototype of the semantic reconstruction method is shown in Fig. 1. This prototype uses $BLIP$ \cite{li2022blip} as the Image2Text model in Fig. 1. The visual encoder of $BLIP$ uses transformer blocks whose weights are initialized from $ViT$-$L/16$ \cite{dosovitskiy2020image}. And the text encoder uses transformer blocks whose weights are initialized from $BERT_{base}$ \cite{devlin2018bert}. For a given text prompt $T$, Stable Diffusion (Text2Image Model in Fig. 1.) \cite{rombach2022high} is used to recover the corresponding image $I$. $Inf(I)$ is calculated by $all$-$MiniLM$-$L6$-$v2$ \cite{wang2020minilm}.

% \begin{table}
% \centering
% \caption{File Size/ Total Persistence  Time Acceleration Times (F.S./TPT AT) compared to the original images.}
% \begin{tabular}{ccccccllll} 
% \hline
%             & TPT AT  & F.S. AT   \\ 
% \hline
% J.75      & 7.06     & 9.99     \\
% J.1       & 42.59    & 74.50    \\
% W.75      & 9.23     & 11.23    \\
% W.1       & 40.46    & 72.81    \\
% Ours      & 163,854.09  & 102,439.75  \\
% Ours*     & 550,349.61  & 432,368.03  \\
% \hline
% \end{tabular}
% \vspace{-1.8em}
% \end{table}

\begin{table*}
\centering
\caption{Comparison of semantic reconstruction with image compression methods. J.75/J.1 means the compressed quality (an integer between 1 and 95) is set to 75/1. J./W. means .jpg/.webp. Ours/ours* means the original text prompt(.txt)/the compressed text prompt(.gz). Definitions of TPT (Total Persistence Time), mWT/MWT (minimum Waiting time/Maximum Waiting Time), F.S. (File Size), Dis. (semantic Distance between the original image and the chosen one), TSS (Time of Semantic Sampling per image)/TSR (Time of Semantic Reconstruction per image) can be found in section 3.1. The best results are in bold.}

\begin{tabular}{ccccccllll} 
\hline
            & TPT(s)  & mWT(ms) & MWT(ms)  & F.S.(MB) & Dis$._{mean}$ & \multicolumn{1}{c}{Dis$._{min}$} & \multicolumn{1}{c}{Dis$._{max}$} & \multicolumn{1}{c}{TSS$_{mean}$(s)} & \multicolumn{1}{c}{TSR$_{mean}$(s)}  \\ 
\hline
Ori.   & \multicolumn{1}{c} {7209.580} & \multicolumn{1}{c}{3744.01} & \multicolumn{1}{c}{19804.14} & \multicolumn{1}{c}{3800.515}              &  \multicolumn{1}{c}{0.0}             & \multicolumn{1}{c}{0.0}                                 &     \multicolumn{1}{c} {0.0}                             &      \multicolumn{1}{c}{-}                             &    \multicolumn{1}{c}{-}                                \\
J.75      & 1020.774 & 533.61  & 2258.87  & 380.160             &     \multicolumn{1}{c}{\textbf{0.253}}           &     \multicolumn{1}{c}{0.0548}  &    \multicolumn{1}{c}{0.763}                   &   \multicolumn{1}{c}{-}  &         \multicolumn{1}{c}{-}    \\
J.1       & 169.245  & 89.48   & 373.03   & 51.007              &   \multicolumn{1}{c}{0.375}             &  \multicolumn{1}{c}{0.115}     &    \multicolumn{1}{c}{0.894}     &       \multicolumn{1}{c}{-}                            &         \multicolumn{1}{c}{-}             \\
W.75      & 780.385  & 457.10  & 1912.74  & 338.183              &    \multicolumn{1}{c}{0.254}            &  \multicolumn{1}{c}{0.0548}    &     \multicolumn{1}{c}{0.700}                              &    \multicolumn{1}{c}{-}                               &          \multicolumn{1}{c}{-}              \\
W.1  & 178.157  & 63.99   & 436.63   & 52.191              &   \multicolumn{1}{c}{0.287}             &          \multicolumn{1}{c}{0.0901}       &  \multicolumn{1}{c}{0.721}           &   \multicolumn{1}{c}{-}                                &          \multicolumn{1}{c}{-}                                                           
\\ S\cite{yang2021slimmable} & 720.143        & 396.13       & 1660.63       & 307.63              &     \multicolumn{1}{c}{0.254}            &  \multicolumn{1}{c}{\textbf{0.0543}}                                &    \multicolumn{1}{c}{\textbf{0.684}}                              &   \multicolumn{1}{c}{-}                                 &   \multicolumn{1}{c}{-}                                 \\

R\cite{mentzer2020learning} & 743.211        &   426.52     & 1820.86        & 324.12              &     \multicolumn{1}{c}{0.254}            &  \multicolumn{1}{c}{0.0544}                                &    \multicolumn{1}{c}{0.697}                              &   \multicolumn{1}{c}{-}                                 &   \multicolumn{1}{c}{-}                                 \\

Ours & 0.0440        & 0.0440       & 0.0440        & 0.0371              &     \multicolumn{1}{c}{0.376}            &  \multicolumn{1}{c}{0.0876}                                &    \multicolumn{1}{c}{0.712}                              &   \multicolumn{1}{c}{12.51}                                 &   \multicolumn{1}{c}{10.11}                                 \\
Ours*  & \textbf{0.0131}       &  \textbf{0.0131}       &  \textbf{0.0131}        & \textbf{0.00879}              &     \multicolumn{1}{c}{0.376}            &  \multicolumn{1}{c}{0.0876}                                &    \multicolumn{1}{c}{0.712}                              &   \multicolumn{1}{c}{12.51}                                 &   \multicolumn{1}{c}{10.11}                              \\
\hline
\end{tabular}
\vspace{-1.8em}
\end{table*}

The results of this experiment are shown in Table 1. We compare our framework with the popular .jpg and .webp methods, state-of-the-art RC  \cite{mentzer2020learning} and  SlimCAEs \cite{yang2021slimmable} methods (PSNR$>$25DB). The W.75/J.75 method in Table 1 means that the original image (.png) in DIV2K is saved as .webp/.jpg using the default quality 75 (PSNR$>$25DB). The W.1/J.1 in Table 1 means that the original image (.png) in DIV2k is saved as .webp/.jpg using quality 1 (PSNR$>$20DB). 
% An example of a high-resolution (2040*1536) image, its compressed results, and its semantic isotope are shown in Fig. 2.  

It is easy to be seen from Table 1 that the framework proposed in this paper has incomparable advantages in persistence speed and file size compared with image compression methods. Because compressing .txt to .gz file does not cause information loss, this method mentioned later in this article refers to the compressed text (.gz file). Even compared with .jpg/.webp, our method can still achieve nearly 10000x in the manner of F.S. AT or TPT AT. mWT and MWT reveal two clues to us. One clue is that for images with similar resolutions, $\frac{MWT}{mWT}$ (for W.1 in Tabel 1) can achieve 6.82 since bpp (bits per pixel) of these images are different. In contrast, the text prompts sampled from the images have similar file sizes. The other clue is that our method can map massive visual data to a file that is small enough to be stored in one blockchain block, which will greatly reduce the communication and synchronization pressures of the entire blockchain system. This explains why the mWT\&MWT metrics corresponding to our method in Table 1 are all consistent with TPT. 

Considering the semantic distance of the two images, we can state that the method proposed in this paper achieves the average semantic distance similar to $Dis_{mean}$ of the .jpg image with quality 1, but only uses up to 1/10000 of the persistence time and 1/6000 of the file size. $Dis_{max}$ reflects the semantic loss in the worst case, and it tells that our method is superior to all the methods in the .jpg series. In other words, this framework can achieve acceptable performance for all images with the worst situation (0.712 in Table 1). From $Dis_{min}$ we can find that this framework can generate extremely accurate semantic expressions in some cases (0.0876 in Table 1).

There will be further doubt that the consumption time of the semantic sampler/reconstructor consumes nearly 10s for a single image, which seems too long. Assuming that there are 1000 hosts with their running nodes on the same blockchain, every node has a different image from DIV2K and they need to persist the images. With our method, there will only be one generated blockchain block and the total time including the TSS\&TSR is nearly 30s. With the J.1 method, the total time will grow up to at least 170s. Besides, for this blockchain, the J.1 file (about 51MB) stored in nearly 510 blockchain blocks has to be verified between 1000 distributed hosts instead of our only local host, and the real total consumption time and the waiting time for each host will increase dramatically with the possible congestion or other factors. The main cause of this phenomenon is that off-chained computing can be parallel while up-chained persistence can only be verified by serialized blocks on the shared blockchain. The effect of TSS\&TSR will decrease with increasing nodes number.

% Besides, we use 4 Nvidia v100 GPUs to run the $BLIP$ and the stable diffusion model. The semantic sampling and the semantic reconstruction will consume average 10s for single image. So the total process of completing a whole process of this framework is nearly 30s. However, these processes can be completed off the blockchain and save the most cherishing file storage and data throughout of the blockchain. 

\begin{table}
\centering
\caption{Semantic distance with/without ROSIS algorithm.}
\begin{tabular}{ccccccllll} 
\hline
            & $Dis_{mean}$  & $Dis_{min}$  & $Dis_{max}$  \\ 
\hline
Ours*      & 0.376     & 0.0876    &  0.712 \\
Worst      & 0.652     & 0.205     &  0.967 \\
Random     & 0.498     & 0.163     &  0.894 \\
\hline
\end{tabular}
\vspace{-1.8em}
\end{table}

\vspace{-1em}
\subsection{Ablation Study of ROSIS Algorithm}
\vspace{-0.45em}

We list the experiment results without using ROSIS in the worst situation and a random situation in Table 2. The worst situation means the semantic isotope is selected as the image whose semantic distance is furthest from the original image and the random situation means the semantic isotope is randomly selected. Compared to the worst situation, the ROSIS algorithm improves by 73.4\% for the mean distance,  134.0\% for the minimum distance, and 35.8\% for the maximum distance. Compared to the random situation, the ROSIS algorithm improves by 32.4\% for the mean distance, 86.0\% for the minimum distance, and 25.6\% for the maximum distance. Although more random seeds will bring better results, the number of random seeds should be considered carefully to save computing resources. 
% \vspace {-1em}
\section{conclusion \& future work}
In the last decade, the bits per pixel increases less than 10\% (vs .webp) on the DIV2K dataset \cite{mentzer2020learning, yang2021slimmable} while the Ethereum users are sharing a network whose total bandwidth is 10MB/min. Therefore, we propose a semantic reconstruction algorithm to build a modal-balanced blockchain under such limited bandwidth. The ROSIS algorithm is designed to reduce the semantic loss caused by the randomness when reconstructing the high dimensional image from the low dimensional text. Experiments prove the breakthrough of making the DIV2K dataset up-chained with less than 1/430,000 space usage and 1/550,000 persistence time with acceptable semantic loss. In the future, we will explore possible algorithms to produce a proper $\hat{T}$ to reduce the difference caused by reconstructor choosing.  Besides, for the dedicated task whose models update at low frequency, we will investigate more time to get a better representation $X$.  

% \vspace {-1em}
\section{acknowledgements}
This work is supported by the Science and Technology Research "$JieBangGuaShuai$" Program of Liaoning Province , China: Intelligent e-Government System based on Consortium Blockchain under Grant 2021JH1/10400010.

\vfill\pagebreak

% \bibliographystyle{IEEEbib}

% \bibliography{strings,refs}

\end{document}